\documentclass[sigconf,screen]{acmart}
\acmConference[ICSE 2026]{IEEE/ACM International Conference on Software Engineering}{12-18 April, 2026}{Rio de Janeiro, Brazil}

\usepackage[normalem]{ulem}
\usepackage{booktabs} 
\usepackage[flushleft]{threeparttable}
\usepackage{xcolor}
\usepackage[utf8]{inputenc}
\usepackage{float}
\usepackage{amsmath}
\usepackage{ifthen}
\usepackage{multirow} 
\usepackage{caption, subcaption}
\usepackage{listings}
\usepackage{url,moreverb,xspace}
\usepackage{tcolorbox}
\usepackage{xspace}

\usepackage{enumitem}
\usepackage{array,graphicx}
\usepackage{soul}
\usepackage{balance}
\usepackage{pifont}
\usepackage{etoolbox,siunitx}
\usepackage{nicefrac}
\usepackage{mathtools}

\usepackage{lipsum}

\usepackage[vlined, boxruled, linesnumbered] {algorithm2e}

\newcommand{\head}[1]{\noindent\textbf{#1.}}

\usepackage{cite}

\usepackage[framemethod=tikz]{mdframed}
\mdfdefinestyle{mpdframe}{
    frametitlebackgroundcolor   =black!15,
    frametitlerule              =true,
    roundcorner                 =3pt,
    middlelinewidth             =1pt,
    innermargin                 =0.6cm, 
    outermargin                 =-0.5cm,
    innerleftmargin             =0.1cm,
    innerrightmargin            =0cm,
    innertopmargin              =0.1cm,
    innerbottommargin           =0.1cm,
    align=center
}

\usepackage{color, colortbl}

\usepackage[nomessages]{fp}

\newcommand{\MyComment}[1]{}

\newcommand{\OurComment}[1]{}
\newcommand{\Space}[1]{}

\definecolor{codered}{rgb}{0.82,0,0}
\definecolor{codeblue}{rgb}{0,0,0.82}
\definecolor{codegreen}{rgb}{0,0.6,0}
\definecolor{codegray}{rgb}{0.5,0.5,0.5}
\definecolor{codepurple}{rgb}{0.58,0,0.82}
\definecolor{lightblue}{rgb}{0.80, 0.90, 1.0}

\newcommand{\changed}[1]{\textcolor{black}{#1}}
\newcommand{\added}[1]{\textcolor{black}{#1}}

\newcommand{\tool}{\textsc{Foresee}\xspace} %
\newcommand{\selforacle}{\textsc{SelfOracle}\xspace} %
\newcommand{\ran}{\textsc{Random}\xspace} 

\newcommand{\greedy}{\textsc{Exhaustive}\xspace}

\newcommand{\tfuse}{\textsc{Transfuser}\xspace}

\newcommand{\ifuse}{\textsc{InterFuser}\xspace}
\newcommand{\apollo}{\textsc{Apollo}\xspace}

\usepackage{booktabs,arydshln}

\makeatletter
\def\adl@drawiv#1#2#3{%
        \hskip.5\tabcolsep
        \xleaders#3{#2.5\@tempdimb #1{1}#2.5\@tempdimb}%
                #2\z@ plus1fil minus1fil\relax
        \hskip.5\tabcolsep}
\newcommand{\cdashlinelr}[1]{%
  \noalign{\vskip\aboverulesep
           \global\let\@dashdrawstore\adl@draw
           \global\let\adl@draw\adl@drawiv}
  \cdashline{#1}
  \noalign{\global\let\adl@draw\@dashdrawstore
           \vskip\belowrulesep}}

\definecolor{Green}{rgb}{0.6,1,0.8}
\definecolor{Red}{rgb}{0.99, 0.76, 0.8}

\definecolor{dkgreen}{rgb}{0,0.6,0}
\definecolor{gray}{rgb}{0.5,0.5,0.5}
\definecolor{mauve}{rgb}{0.58,0,0.82}

\newcommand{\ego}{ego vehicle\xspace}

\newcommand{\totalFailuresForesee}{609} 

\newcommand{\infractionPercentage}{\textsc{6.74}} 

\newcommand{\FOImprovementOverSOIfuse}{\textsc{44.09}} 

\newcommand{\FOImprovementOverRandomIfuse}{\textsc{186.69}}

\newcommand{\FOImprovementOverSOTfuse}{\textsc{21.38}} 

\newcommand{\FOImprovementOverRandomTfuse}{\textsc{37.06}}

\newcommand{\FOImprovementOverSO}{\textsc{38.09}} 

\newcommand{\FOImprovementOverRandom}{\textsc{128.70}}

\newcommand{\configCount}{\textsc{12}\xspace}

\newcommand{\numOfTechs}{3\xspace} 
\newcommand{\mutationCount}{4\xspace} 
\newcommand{\maxNrp}{4\xspace} 
\newcommand{\repetitionCount}{\textsc{3}\xspace} 
\newcommand{\scenarioCount}{\textsc{5}\xspace} 
\newcommand{\totalRouteCount}{\textsc{120}\xspace} 

\newcommand{\routesWithInfractionsIf}{\textsc{17}\xspace}
\newcommand{\routesWithInfractionsTr}{\textsc{9}\xspace}

\newcommand{\remainingRoutesIf}{\textsc{103}\xspace}
\newcommand{\remainingRoutesTr}{\textsc{111}\xspace}

\newcommand{\totalClippedIf}{297}
\newcommand{\totalClippedTr}{181}

\newcommand{\runWithRepIf}{1,782}    
\newcommand{\runWithRepTr}{1,086}    

\newcommand{\runTotal}{\textsc{34,416}\xspace} 

\newcommand{\avgDuration}{\textsc{22.5}}

\newcommand{\avgDurationSixs}{\textsc{15}}

\newcommand{\avgDurationTens}{\textsc{30}}

\newcommand{\totalDuration}{\textsc{215.1}}

\newcommand{\greedyDuration}{\textsc{43}}

\newcommand{\foreseeAUCIf}{\textsc{1910171.65}} 

\newcommand{\foreseeBySelfOracleAUCIf}{\textsc{1.44$\times$}} 
\newcommand{\foreseeByRandomAUCIf}{\textsc{2.62$\times$}} 

\newcommand{\foreseeAUCTr}{\textsc{223400.39}} 

\newcommand{\foreseeBySelfOracleAUCTr}{\textsc{1.27$\times$}} 
\newcommand{\foreseeByRandomAUCTr}{\textsc{1.74$\times$}} 

\newcommand{\foreseeBySelfOracleAUC}{\textsc{1.42$\times$}} 
\newcommand{\foreseeByRandomAUC}{\textsc{2.49$\times$}} 

\newcommand{\chosenDuration}{\textsc{10}} 
\newcommand{\chosenNrp}{\textsc{4}} 

\newcommand{\rpSixOne}{\textsc{91}} 
\newcommand{\colSixOne}{\textsc{16.33}} 
\newcommand{\rateSixOne}{\textsc{17.95}} 

\newcommand{\rpSixTwo}{\textsc{179.33}} 
\newcommand{\colSixTwo}{\textsc{41}} 
\newcommand{\rateSixTwo}{\textsc{22.86}} 

\newcommand{\rpSixFour}{\textsc{275.33}} 
\newcommand{\colSixFour}{\textsc{73}} 
\newcommand{\rateSixFour}{\textsc{26.51}} 

\newcommand{\rpTenOne}{\textsc{88}} 
\newcommand{\colTenOne}{\textsc{24}} 
\newcommand{\rateTenOne}{\textsc{27.27}} 

\newcommand{\rpTenTwo}{\textsc{174}} 
\newcommand{\colTenTwo}{\textsc{57}} 
\newcommand{\rateTenTwo}{\textsc{32.76}} 

\newcommand{\rpTenFour}{\textsc{274}} 
\newcommand{\colTenFour}{\textsc{93.67}} 
\newcommand{\rateTenFour}{\textsc{34.19}} 

\newcommand{\avgDurationSix}{\textsc{4.7h}}
\newcommand{\avgDurationTen}{\textsc{6.4h}}

\newcommand{\greedySix}{\textsc{271}} 
\newcommand{\greedySixSimTime}{\textsc{64h}} 
\newcommand{\greedySixColByHour}{\textsc{4.2}} 

\newcommand{\coverageSix}{\textsc{26.94}} 
\newcommand{\greedyTen}{\textsc{233.33}}
\newcommand{\greedyTenSimTime}{\textsc{66.5h}} 
\newcommand{\greedyTenColByHour}{\textsc{3.5}} 

\newcommand{\coverageTen}{\textsc{40.14}} 

\newcommand{\foreseeSixColByHour}{\textsc{15.53}} 
\newcommand{\foreseeTenColByHour}{\textsc{14.64}} 

\newcommand{\efficiencySixPercent}{\textsc{269.76}}
\newcommand{\efficiencyTenPercent}{\textsc{318.29}}

\newcommand{\TS}{$\mathit{TS}$\xspace}

\renewcommand{\cite}{\citep}

\setcopyright{acmcopyright}

\begin{document}

\title{Misbehavior Forecasting for Focused Autonomous Driving Systems Testing}

\author{M M Abid Naziri}
\orcid{0000-0002-3499-5283}
\email{mnaziri@ncsu.edu}
\authornotemark[1]
\affiliation{%
  \institution{North Carolina State University}
  \streetaddress{2101 Hillsborough Street}
  \city{Raleigh}
  \state{North Carolina}
  \country{USA}
  \postcode{27607}
}

\author{Stefano Carlo Lambertenghi}
\orcid{0009-0007-0513-1877}
\email{lambertenghi@fortiss.org}
\authornotemark[3]
\affiliation{%
  \institution{Technical University of Munich, fortiss GmbH}
  \streetaddress{Guerickestra{\ss}e 25}
  \city{Munich}
  \state{Bayern}
  \country{Germany}
  \postcode{80805}
}

\author{Andrea Stocco}
\orcid{0000-0001-8956-3894}
\authornotemark[2]
\email{andrea.stocco@tum.de}
\affiliation{%
  \institution{Technical University of Munich, fortiss GmbH}
  \streetaddress{Boltzmannstra{\ss}e 3}
  \city{Munich}
  \state{Bayern}
  \country{Germany}
  \postcode{85748}
}
\authornotemark[3]

\author{Marcelo d'Amorim}
\orcid{0000-0002-1323-8769}
\email{mdamori@ncsu.edu}
\authornotemark[1]
\affiliation{%
  \institution{North Carolina State University}
  \streetaddress{2101 Hillsborough Street}
  \city{Raleigh}
  \state{North Carolina}
  \country{USA}
  \postcode{27607}
}

\renewcommand{\shortauthors}{Naziri et al.}

\begin{abstract}
Simulation-based testing is the standard practice for assessing the reliability of self-driving cars' software before deployment. 
Existing bug-finding techniques are either unreliable or expensive.
We build on the insight that near misses observed during simulations may point to potential failures. We propose \tool, a technique that identifies near misses using a misbehavior forecaster that computes possible future states of the ego-vehicle under test.
\tool performs local fuzzing in the neighborhood of each candidate near miss to surface previously unknown failures.
In our empirical study, we evaluate the effectiveness of different configurations of \tool using several scenarios provided in the CARLA simulator on both end-to-end and modular self-driving systems and examine its complementarity with the state-of-the-art fuzzer DriveFuzz.
Our results show that \tool is both more effective and more efficient than the baselines. \tool exposes \FOImprovementOverRandom\% and \FOImprovementOverSO\% more failures than a random approach and a state-of-the-art failure predictor while being \foreseeByRandomAUC~and \foreseeBySelfOracleAUC~faster, respectively. Moreover, when used in combination with DriveFuzz, \tool enhances failure detection by up to 93.94\%.

\end{abstract}

\maketitle

\section{Introduction}\label{sec:introduction}

Simulation-based testing~\cite{kaur2021survey,Guannam_ETAL_FSE22} is the de facto approach for Autonomous driving systems (ADS) testing~\cite{ADS_testing_survey_2022,zhong2021surveyscenariobasedtestingautomated} as real-world physical testing, albeit important, has severe time, resource, and legal limitations~\cite{2022-Stocco-TSE}. Simulators enable developers to assess the reliability of the ADS before deployment as they consist of virtual simulation platforms in which developers ``plug'' their ADS and test it against challenging conditions. A test describes a route that an ADS must complete within a map representing an urban environment containing static and dynamic objects (e.g., traffic signs and other vehicles). 
Prior work has been proposed to generate test cases for ADS, particularly leveraging search-based optimization~\cite{2020-Riccio-FSE,Abdessalem-ASE18-1,Abdessalem-ASE18-2,Abdessalem-ICSE18,9712397,Guannam_ETAL_FSE22,MULLINS2018197,Gambi:2019:ATS:3293882.3330566} and fuzzing~\cite{drivefuzz,zhongETAL2021,liETAL2020,Jha2019MLBasedFI,10.1145/3597926.3598072}, which are characterized by drawbacks in terms of effectiveness and efficiency. Concerning the former, these solutions require the exploration of a vast multi-dimensional search space to pinpoint critical conditions. This is problematic due to the significant time and resource overhead, as running a single test case on a driving simulation platform can take several minutes. About the latter, existing testing techniques either apply mutations \emph{at the scenario level} in the initial state of a simulation, e.g., the number and position of the vehicles at the beginning of the simulation, or guide the search relying on the \emph{global} ADS behavior during the entire simulation or static features of test cases, such as the number of bends of a road~\cite{2024-Biagiola-EMSE}.
As a consequence, these techniques are oblivious to \emph{near misses}, i.e., circumstances where a failure would occur with minimal modifications to certain intermediate states of a simulation in which the ADS was close to failure.
Overlooking near misses limits the potential of test generators to reveal failures and potentially inflates the safety perception of ADS.

This paper aims to improve simulation-based testing of ADS in terms of effectiveness and efficiency by investigating the problem of \emph{detecting and exploiting} near misses during virtual simulations. 
Instead of relying on global or initial states of a scenario-based test, we target focused testing of intermediate states occurring during failure-free simulations. We leverage the insight that critical scenarios may exist within driving simulations, where even minor modifications, such as slight speed adjustments at an intersection, could expose failures. 

\begin{figure}[t]
\centering     
\includegraphics[width=0.9\linewidth]{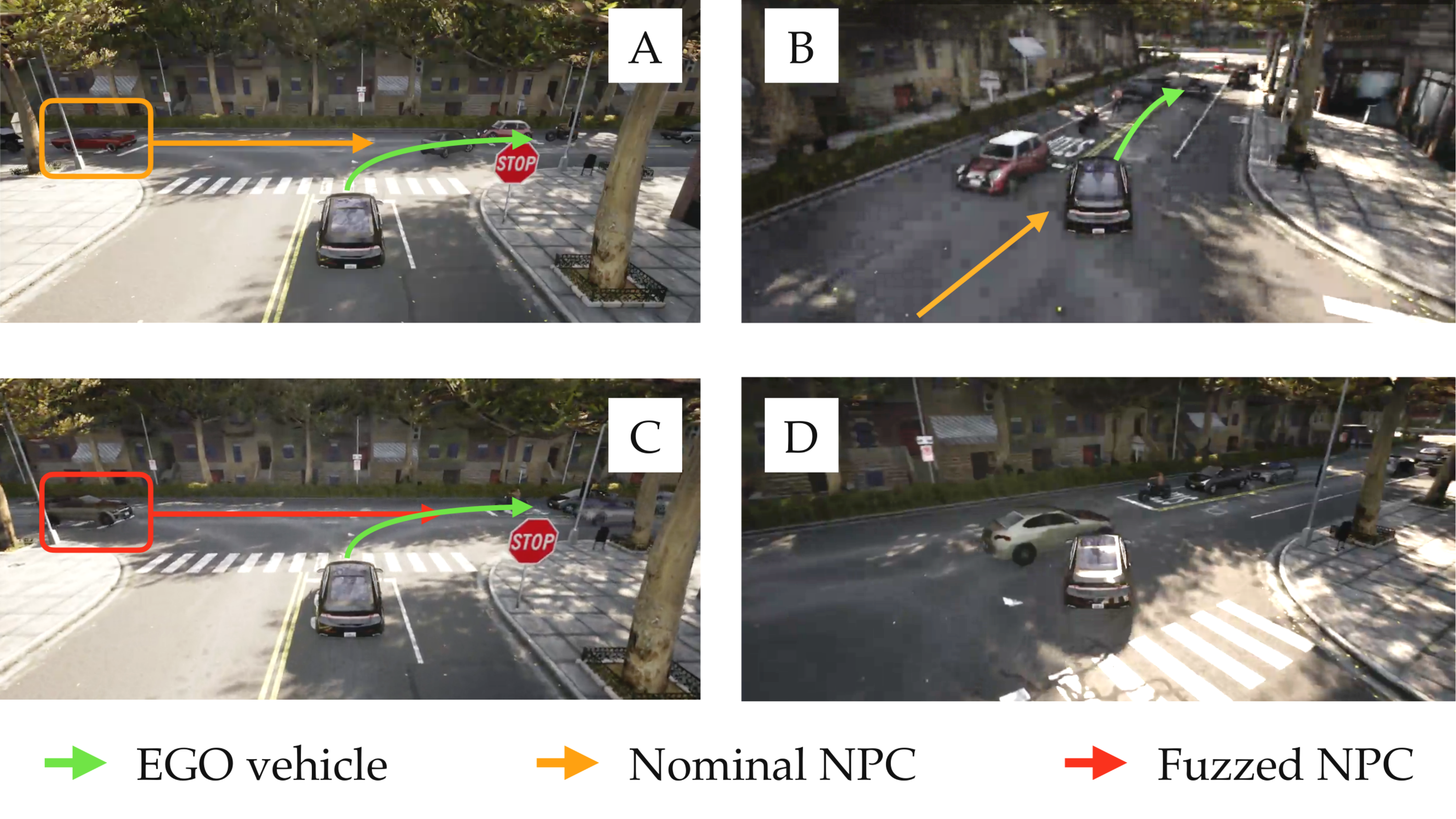}
\caption{Illustrative example showing a near miss from a failure-free simulation (A/B) and collision from a sub-simulation which was mutated and clipped by \tool (C/D).}
\label{fig:visual_example}
\end{figure}

We propose \tool (\underline{FORE}casting un\underline{S}afe \underline{E}vents and \underline{E}mergency situations), a \emph{focused} system testing technique for ADS. \tool uses a monitor to measure risk during the simulation of a given test case. It uses risk data to derive, classify, and prioritize short-running test cases. \tool fuzzes the inputs of these local and seemingly relevant test cases to find failures. More specifically, \tool uses telemetry data (i.e., velocity, yaw rate, position, and relative distances to other agents) collected during simulation to automatically identify conditions in which the system is close to failure.
This paper focuses on the identification of collisions, being the primary acceptance criteria for the safe deployment of ADS. We show that telemetry data offers clues about the failure likelihood of the ADS. \tool forecasts potential failing conditions, such as the identification of vehicles or pedestrians that are crossing the future trajectory of the vehicle under test. 
Hazardous driving conditions are detected when the failure likelihood increases in a future state of the simulation, as predicted by \tool. 

\autoref{fig:visual_example} shows an illustrative example of a near-miss situation and \tool in action. The images \textbf{A} and \textbf{B} display consecutive snapshots of a nominal failure-free simulation. Note that the (future) trajectory of the \ego and the trajectory of the NPC vehicle would eventually cross. However, due to the low speed of the NPC, the two vehicles do \emph{not} collide (image~\textbf{B}). We refer to such missed cases from the original simulations as ``near misses''. Images \textbf{C} and \textbf{D} display a failure condition that \tool reports. \tool identifies the simulation segment at which the vehicles would have eventually intersected as a risky point. Then, it  ``clips'' the risky segment from the original simulation and creates new focused simulations derived from the clipped segment. In this example, the new simulation is obtained by mutating the model of the NPC vehicle (image \textbf{C}). The new simulation results in a collision between the \ego and the mutated NPC, which is faster than the original NPC (image \textbf{D}).

We evaluated the effectiveness of \tool in the CARLA simulator~\cite{carla}, using ADS available from the literature and a diverse set of complex urban scenarios in which we observed many near-miss situations~\cite{carla-scenarios}. 
In our experiments on \changed{\runTotal} simulations accounting for more than \totalFailuresForesee~individual failures, \tool was able to surface up to \infractionPercentage\% additional failures from near misses, a \FOImprovementOverSO\% increase with respect to \selforacle~\cite{2020-Stocco-ICSE}, a state-of-the-art misbehavior predictor based on autoencoders, and a \FOImprovementOverRandom\% increase with respect to a random assessment of risk. \tool also demonstrates its efficiency against all the other baselines by discovering collisions \foreseeBySelfOracleAUC~faster than \selforacle and \foreseeByRandomAUC~faster than \ran. We also observed a \efficiencyTenPercent \% more efficient discovery of failures with the best-performing configuration of \tool relative to a ground truth generated from an exhaustive search. Additionally, we show that \tool complements the state-of-the-art fuzzer DriveFuzz up to a 93.94\% failure exposure increase.

Our paper makes the following contributions:

\head{Technique}
A technique for forecasting ADS misbehavior based on the kinematic state of the ego vehicle. Our approach is implemented in the publicly available tool \tool~\cite{replication-package}.

\head{Evaluation}
An empirical study showing that \tool's risk-based assessment outperforms a random and a black-box approach~\selforacle~\cite{2020-Stocco-ICSE} in terms of near-misses detection and improves the fuzzer DriveFuzz~\cite{drivefuzz} in terms of failure exposure.

\head{Dataset}
A dataset of ADS failures to evaluate the performance of failure prediction systems and test generators for ADS. The tool and evaluation data are publicly available~\cite{replication-package}. 

\section{Background}\label{sec:background}

ADS are software systems developed with increasing capabilities to drive vehicles autonomously. ADS are usually designed following a modular architecture that includes perception, prediction, planning, and control modules~\cite{Kato_ETAL_ICCPS2018,yurtsever2020survey,apollo}. 
Driving simulation platforms are the de facto choice in the industry for developing and testing ADS before real-world testing on roads~\cite{2022-Stocco-TSE,2020-Haq-ICST,2021-Haq-EMSE}. 
In the remainder of this section, we describe the nomenclature used in this paper.

\head{Scenario} 
Scenarios define high-level traffic situations involving vehicle movements and interactions, commonly used for testing ADS behavior. These are often derived from real-world data, such as “pre-crash” reports from agencies like the US NHTSA~\cite{us-drive-safety-facts2017}, and are used to evaluate safety-critical responses.
A scenario specifies the sequence of events in a simulation, including interactions with actors such as pedestrians jaywalking, or vehicles running red lights. It incorporates both static elements (e.g., traffic lights, crosswalks, trees) defined by the map layout, and dynamic elements such as the \ego and Non-Playable Characters (NPCs), which follow scripted behaviors.
Scenarios help uncover failures by capturing both routine and unexpected interactions between vehicles, infrastructure, and pedestrians.

\head{Test Case}
Given a scenario, a simulation-based test case is characterized by an \emph{initial state} and a \emph{route}. The initial state outlines the conditions of both static and dynamic objects at the beginning of the simulation, including the positions, velocities, and states of all objects in the scenario. The route specifies the path the \ego is expected to follow during the simulation. It is typically defined in terms of a starting and ending point or as a sequence of waypoints within the map through which the \ego\ should navigate. In summary, a route establishes a possible ground truth trajectory that the \ego should follow in the simulation. 

\head{Failures}
ADS are designed to meet several requirements, encompassing factors about passenger safety and comfort~\cite{10.1007/978-3-658-21194-3_53}. Driving simulation platforms automatically log any rule violations that occur during testing. Among these, safety violations are of utmost concern, particularly when it comes to autonomous driving, as they can potentially lead to vehicle crashes and casualties. 
This work focuses on predicting collisions.
Collision avoidance is a primary prerequisite to be met as a self-driving vehicle must stay in its lane and prevent collisions to gain public trust and acceptance for production use. 
Our categories of failures include collisions involving the \ego with elements beyond the road, such as pavements or poles, pedestrians, or other vehicles (\autoref{sec:setup}).

\section{Approach}\label{sec:approach}

\begin{figure*}[t]
\centering     
\includegraphics[trim={0.4cm 19.2cm 0.6cm 0.0cm}, clip, width=0.92\textwidth]{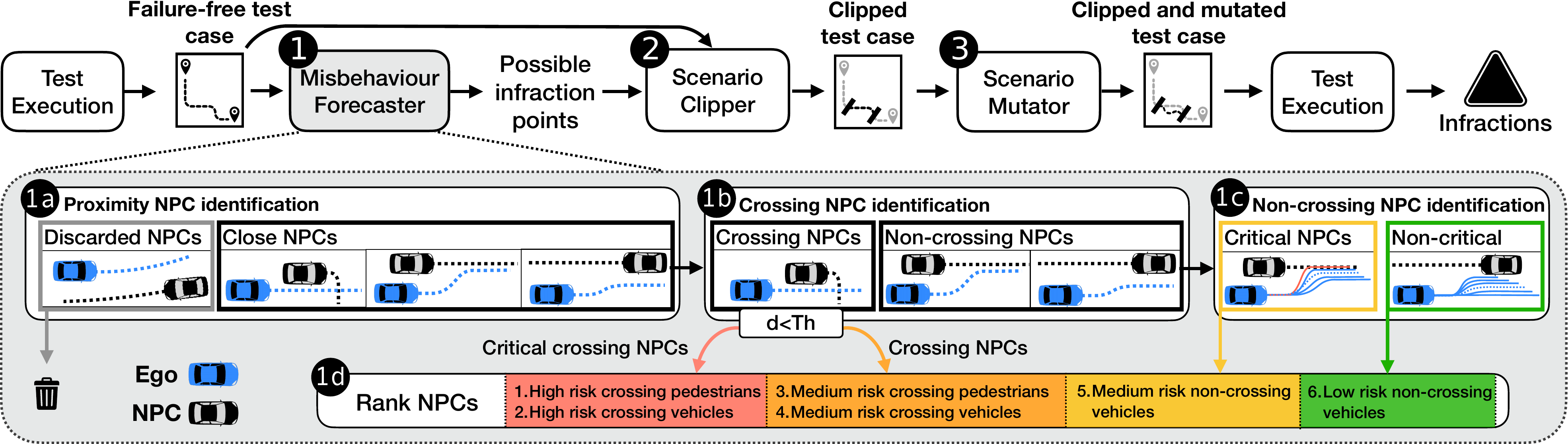}
\caption{An overview of \tool (top), and the logic of the Misbehavior Forecaster in detail (bottom).}
\label{fig:pipeline}
\end{figure*}

\tool aims to detect the occurrence of unexposed system failures during simulation-based testing of ADS. 
It builds on the observation that infractions are relatively rare compared to near misses. For example, on average, DriveFuzz~\cite{drivefuzz} exposed 19 violations in 360 minutes ($\approx$19 minutes per failure), whereas AV-Fuzzer~\cite{liETAL2020} exposed, on average, 50 failures in 1,000 simulations (20 simulations per failure)~\cite{zhongETAL2021}.
The unique aspect of \tool is that \emph{it exploits near misses observed during simulations} to detect failures. 

A test suite \TS consists of test cases associated with scenarios that are challenging for an autonomous vehicle. For example, Scenario~4 
of the CARLA leaderboard~\cite{carla-scenarios} deals with situations where the \ego finds an obstacle on the road while performing a maneuver, and it must perform an emergency brake or an avoidance maneuver. 

\tool focuses on failure-free test cases, aiming to uncover missed failures that occur in near-critical conditions. This focus is motivated by the high cost of simulation-based testing, which makes it valuable to reuse simulations that already exhibit such near-critical behavior.
Although any test could theoretically be forced into failure with unrealistic changes (e.g., excessive NPC speeds), \tool prioritizes preserving realism. It introduces only small, targeted mutations and applies sanity checks to ensure consistency with the original scenario’s intent.

\subsection{Overview}

\tool takes as input a test case that does not reveal failures in nominal conditions (i.e., a failure-free test case) and reports infractions as output. The upper portion of \autoref{fig:pipeline} illustrates the \tool pipeline, consisting of three tasks, namely Misbehavior Forecaster, Scenario Clipper, and Scenario Mutator. 

The Misbehavior Forecaster~\ding{182} is responsible for identifying risky conditions during a simulation by tracking the position, speed, and steering angle of the NPCs at each time frame. Using the information in the collected execution traces, it then predicts which parts of the simulation are more likely to cause infractions.
This component reports a ranked list of simulation timestamps sorted by a criticality score based on the probability of the \ego to intersect NPC future trajectories; we hereafter refer to this list's elements as risky points.
\tool attempts to find potential modifications in the original simulation around these risky points to reveal previously unforeseen failures. 
To this aim, the Scenario Clipper~\ding{183} is used to reconstruct a feasible CARLA-runnable scenario in the neighborhood of a risky point. For each risky point, \tool computes start and end points from the original simulation that include the given likely infraction-inducing point. Then, it retrieves the list of NPCs relevant for that subset of the simulation. This is done to reduce unnecessary computation and focus risk assessment on meaningful threat sources.
Finally, the Scenario Mutator~\ding{184} introduces mutations in the initial states of the new scenario, effectively creating small variations on the intermediate states of the original simulation. At the end of the process, \tool runs the derived (short) simulations and reports test cases revealing infractions. 

In summary, \tool uses a combination of misbehavior forecasting, scenario clipping, and local scenario-level mutation to find unforeseen failures. It reports failure-revealing test cases without executing expensive fuzzing campaigns.
 
\subsection{Misbehavior Forecaster}\label{sec:misbehavior-forecaster}

In this paper, we propose a novel type of predictor in which risk is assessed using telemetry data to forecast potential failing conditions, such as vehicles or pedestrians that are crossing the future trajectory of the vehicle under test. 

The misbehavior forecaster takes a test case $tc$ as input and returns a list of risky points ranked by their likelihood of causing an infraction. 
The lower part of \autoref{fig:pipeline} illustrates the four-step workflow of \tool to obtain the ranked list of NPCs.

\head{Step 1a: Proximity NPC identification}
This step identifies the NPCs that approach the \ego within a certain radius during the original simulation.
For that, \tool computes the set \textit{Close NPCs} describing the circumstance when NPCs are closest to the \ego, considering that radius. The set contains pairs with the simulation frame of the close encounter and the simulation ID of the corresponding NPC, and is further categorised in subsequent steps.
Any NPCs that are not within the threshold radius are not considered (\textit{Discarded NPCs}).

\head{Step 1b: Crossing NPC identification}
For each NPC identified in the previous step, \tool retains those that cross the \ego trajectory for further categorization. This filtering process yields two groups of NPCs: \textit{Crossing NPCs} contains NPCs that intersect with the \ego path during any simulation frame, and its subset, \textit{Critical crossing NPCs}, includes NPCs that cross the \ego but only within a limited distance from it. The remaining NPCs are identified as \textit{Non crossing NPCs}$=\{ x \, | \, x \in \text{\textit{Close NPCs}} \text{ and } x \notin \text{\textit{Crossing NPCs}} \}$ and are further analyzed in step 1c.

\head{Step 1c: Non-crossing NPC identification} 
\tool uses this step to identify vehicle trajectories that currently do not cause collisions, but could result in collisions with the \ego with minimal changes, such as slight modifications of NPC behaviour (e.g., speed, or steering profiles).
For each NPC in set \textit{Non crossing NPCs}, \tool generates $N$ perturbations of the original \ego trajectory by introducing a small error in the velocity and yaw rate values, ensuring that the perturbations remain within acceptable localization error bounds, following the existing guidelines~\cite{localization_req}.

It then evaluates if any of the newly generated \ego trajectories is below a distance $threshold$ to the NPC. If so, \tool saves the NPC in the set \added{\textit{Critical NPCs}}.
All \textit{Non-crossing NPCs} ids that do not meet this distance requirement are saved in \textit{Non-critical NPCs}.
The goal of this procedure is to account for the randomicity of the self-driving system under test, which could behave in a slightly different manner between different runs of the same scenario, therefore producing an unforeseen, dangerous situation. 

\head{Step 1d: Rank NPCs} 
In this step, \tool ranks the risky points associated with the NPCs in the aforementioned groups according to their risk of causing a collision with the \ego, and based on the type of NPC: pedestrians or vehicles (see step 1d of \autoref{fig:pipeline}). If multiple NPCs are associated with the same risk level, \tool gives a higher score to NPCs that come closer to the \ego trajectory during the simulation. 
For each of the ranked NPCs, our approach collects the simulation $\mathit{frame}$ at which the actor comes closest to the \ego during the nominal simulation. 

\subsection{Scenario Clipper}\label{sec:scenario-clipper}

The scenario clipper component of \tool is responsible for creating a scenario reflecting solely the risky conditions observed during the execution of the original scenario. We use the term ``clip'' to indicate that only a subset of the original scenario is retained.

The top $\mathit{n_{rp}}$ risky points identified by the Misbehavior Forecaster (\autoref{sec:misbehavior-forecaster}) determine the timestamps at which the simulation will be clipped, where $\mathit{n_{rp}}$ is a hyperparameter that determines how many risky points should be considered for clipping.
The parameters $o_b$ and $o_a$ indicate the length of the clip.
For each chosen risky point $\mathit{rp}$, \tool clips the scenario from timestamp $\mathit{rp - o_b}$ to timestamp $\mathit{rp + o_a}$. To restore the state of the original simulation at timestamp $rp - o_b$, \tool saves the location, direction, and model of each NPC. In this way, the clipped scenario is ``centered around'' the risky point $\mathit{rp}$. 
For timestamp $\mathit{rp - o_b}$, our approach stores the exact location of the \ego at that timestamp. This information is useful to set the starting waypoint $\mathit{s_{wp}}$ for the clipped scenario. 
Concerning the ending waypoint $\mathit{e_{wp}}$, the selection is more challenging. Indeed, we observed that the location of the \ego at timestamp $\mathit{rp + o_a}$ often results in invalid simulations in CARLA, because the simulator maintains specific sets of waypoints that can be used as a route in a scenario. As a working solution, our approach retrieves, from the log of the original simulation, a list of valid waypoints and uses the closest waypoint to the location of the \ego at timestamp $\mathit{rp + o_a}$ as the ending waypoint $\mathit{e_{wp}}$. Finally, a new route XML file \changed{is created} in which the clipped scenario starts at $\mathit{s_{wp}}$ and ends at $\mathit{e_{wp}}$. 

\subsection{Scenario Mutator}\label{sec:scenario-mutator}

For each clipped scenario, \tool applies NPC-focused mutations. 
The rationale is to assess whether the \ego can cope with situations that are \emph{analogous} to the one observed during the riskiest parts of the original simulation, yet they are slightly different. For each risky point, \tool generates $c$ mutated children (\autoref{sec:scenario-clipper}), executes them, and reports the number of collisions.
Our approach retains the original number of NPCs while varying certain properties. Two mutation operators are currently supported.

\head{NPC Model Swapping} The Scenario Mutator swaps an NPC's vehicle model with another of the same type. For example, a bicycle may be replaced with another bicycle or a pedestrian, while a car is only substituted with another car. This can impact the vehicle's kinematic characteristics, such as speed differences between the original and replacement car. The replacements are limited to the ``relevant'' NPCs in the neighborhood of the \ego. To identify these relevant NPCs, the misbehavior forecaster reports a ranked list of NPCs that approached the \ego along with the time frame in which they were closest. Of these, only the closest is selected for mutation.
\tool uses this information to select the NPCs for which their time frame lies within the interval $\mathit{[rp - o_b, rp + o_a]}$, where $rp$ denotes the riskiest point selected from the ranked list and $\mathit{o_b}$ and $\mathit{o_a}$ denote, respectively, the offsets before and after the risky point delimiting the period of a new simulation.

\head{Steering Angle Perturbation} The Scenario Mutator perturbs the current steering angle of the closest NPC to the \ego as another form of mutation. To introduce variations in the steering angle, \tool tracks the closest vehicle to the \ego $\mathit{npc_{closest}}$ in a test case at the beginning of the risky interval, i.e., at $\mathit{rp - o_b}$. Afterwards, for each child simulation, the Scenario Mutator applies a random steering angle to $npc_{closest}$. The simulator accepts values within the range of $[-1.0, 1.0]$; a value within this range is randomly selected for the steering angle, following existing thresholds~\cite{localization_req}. 

\head{Validity/Realism Check}
To maintain the validity and realism of the original simulation, the Scenario Mutator produces new short-lived simulations introducing modifications in existing NPCs within the domain model and constraints of the CARLA simulator. This ensures that the resulting mutations are valid and realistic by design, as they operate within the NPC and kinematic space allowed in the CARLA simulator.
The Scenario Mutator also ensures that the newly mutated vehicle model avoids collisions with other NPC vehicles upon spawning at the beginning of the simulation, potentially due to the increased length of the new model vehicle.
To achieve this, our approach computes the distances between each pair of NPC vehicles and retains only the valid vehicle models. A valid vehicle model fits within the gap between two adjacent NPCs and does not cause immediate collisions. 
Additionally, certain sensors on the map at the initial point of a simulation can be placed as invisible objects, causing collisions with NPC vehicles if placed directly on the road. To avoid inflating the number of collisions with these phantom objects, we increment the z-axis value of the locations by a small constant value $\mathit{z\_offset}=2$ when saving the location. Since the simulator accounts for gravity, these vehicles are automatically positioned on the ground upon spawning.

\section{Evaluation}
\label{sec:empirical-study}

\subsection{Research Questions}\label{sec:research-questions}

\textbf{RQ\textsubscript{1} (effectiveness):}
\newcommand{\rqEffectiveness}{How effective is \tool in exposing misbehaviors in near-miss scenarios compared to exhaustive search? How does effectiveness vary with clip sizes and number of child tests?}
\textit{\rqEffectiveness}

\noindent
\textbf{RQ\textsubscript{2} (comparison):}
\newcommand{\rqCompare}{How does \tool compare with alternative misbehavior prediction techniques (\greedy, \ran and \selforacle) and proximity-based ranking?}
\textit{\rqCompare}

\noindent
\textbf{RQ\textsubscript{3} (efficiency):}
\newcommand{\rqEfficiency}{How efficient is \tool in exposing misbehaviors in near-critical situations?}
\textit{\rqEfficiency}

\noindent
\textbf{RQ\textsubscript{4} (complementarity):}
\newcommand{\rqFaultRevealing}{Does \tool improve an existing state-of-the-art fuzzer?}
\textit{\rqFaultRevealing}

\noindent
\changed{\textbf{RQ\textsubscript{5} (generalizability):}}
\newcommand{\rqGeneralizing}{Does \tool generalize to other scenarios and industry-grade ADS?}
\changed{\textit{\rqGeneralizing}}

The first research question evaluates the ability of \tool to detect near misses and generate failures. We evaluate effectiveness when varying two important parameters: the size of the clipped scenarios (as per $o_a$+$o_b$) and the number of near misses to exploit in each test (as per $n_{rp}$). Intuitively, longer clips and a higher number of risky points may reduce \ the tool's efficacy as the technique is tailored for targeted risk selection. We also compare the precision of \tool against an impractical \greedy approach that approximates an upper bound on the possible number of failures by applying fuzzing at each timestamp (excluding the first $o_b$ seconds). 

The goal of the second research question is to measure the ability of \tool's misbehavior forecaster component~(Section~\ref{sec:misbehavior-forecaster}) to detect risky points. We compare \tool against two baseline approaches, namely \ran and \selforacle. The \ran baseline selects a waypoint for local fuzzing at random, clips the simulation around that waypoint, and mutates its corresponding initial state. The second approach replaces the kinetics-driven misbehavior forecaster of \tool with \selforacle~\cite{2020-Stocco-ICSE}, a data-driven misbehavior predictor for ADS based on autoencoders.

The third research question evaluates how fast \tool exposes misbehaviors. 
We hypothesize that \tool exposes failures faster because only parts of the original simulation are used and executed. This research question evaluates this hypothesis.

The fourth research question evaluates whether \tool can complement an existing state-of-the-art fuzzer, specifically DriveFuzz~\cite{drivefuzz}, by uncovering failures in non-failing test cases.

\added{Lastly, the fifth research question evaluates the generalizability of \tool across additional complex urban scenarios, and an evaluation using an industrial-grade ADS.}

\subsection{Objects: Simulator, Scenarios, and ADS}\label{sec:scenarios}

\subsubsection{Simulator}

We used the CARLA simulator for self-driving cars~\cite{carla} (v. 0.9.10.1), a driving simulator developed with the Unreal Engine 4~\cite{unreal} used in previous ADS testing literature~\cite{zhongETAL2021,drivefuzz,9712397}. We chose CARLA as it supports complex urban scenarios with many configurations of static and dynamic objects, and provides a rich set of sensors (e.g., cameras, LiDAR, GPS, and radar) to enable the observation of the status of the ADS throughout the simulation. 

\subsubsection{Scenarios}

CARLA provides multiple closed-loop urban maps for ADS testing. We use Town10, a default map with standard environmental settings (e.g., sunny weather). Each map includes driving scenarios, defined by multiple test cases (called routes in CARLA). \autoref{table:scenario-characterization} summarizes the five selected scenarios, chosen for their diversity, totaling 120 test cases.

\begin{table}[t]
    \caption{\label{table:scenario-characterization}Characterization of scenarios of Town10.}
    \label{table:scenario-characterization} 
    \centering
    \footnotesize
    \setlength{\tabcolsep}{4.1pt}
    \renewcommand{\arraystretch}{1}
        \begin{tabular}{clc} 
        \toprule
        \textbf{Scenario} & \textbf{Description} & \textbf{\# Routes} \\
        \midrule
        3 & NPCs cross in front of the \ego, requiring braking & 16 \\
        4 & Obstacles appear post-turn, requiring reactive avoidance & 46 \\
        7 & NPCs run red lights; \ego must avoid collisions & 19 \\
        8 & \ego makes unprotected left turn, yields to traffic & 19 \\
        9 & \ego turns right, yielding to crossing traffic & 20 \\
        \bottomrule
        \end{tabular}
\end{table}

\subsubsection{ADS under Test}

We use \ifuse~\cite{shao2023safety} and \tfuse~\cite{Chitta2022PAMI}, two ADS that achieve top performance on the CARLA leaderboard~\cite{carla-leaderboard} and have been adopted in recent work~\cite{jia2023think,jia2023driveadapter,jaeger2023hidden}.

\ifuse is a multi-modal fusion model designed for complex driving scenarios. It uses camera and LiDAR inputs, along with interpretable intermediate features (e.g., planned trajectory, traffic signals), to produce safe driving commands (e.g., steering, throttle). 
Features are extracted using ResNet backbones and fused via a transformer module, followed by a Gated Recurrent Unit for trajectory prediction. A safety controller uses high-level cues to constrain final control outputs.

\tfuse is a transformer-based fusion model that integrates spatial and temporal features from camera and LiDAR inputs using cross-attention~\cite{NIPS2017_3f5ee243}. Unlike \ifuse, \tfuse performs sensor fusion at a later stage, combining image and LiDAR features extracted via ResNet encoders. 
A waypoint prediction module outputs the \ego vehicle's future trajectory, which informs low-level driving actions.

\subsection{Competing Methods}\label{sec:baselines}

Since direct baselines for our approach are not available, we conduct experiments exploring different methods for detecting risky points and applying alternative ranking strategies, as described below.

\subsubsection{Risk Scenario Identification}
To evaluate risk-scenario identification, we compare \tool against three baseline variants that retain the Scenario Clipper~(\autoref{sec:scenario-clipper}) and Scenario Mutator~(\autoref{sec:scenario-mutator}) components, but substitute the misbehavior forecaster with alternative strategies, described below.

\head{\greedy}\label{sec:baseline-greedy} 
We compute an upper bound on failure detection by exhaustively sampling the simulation timeline, independent of risk scores. Although computationally expensive and impractical in real-world use, it serves as a reference for maximum possible coverage.
\greedy selects every second of the original simulation as the center of a clip, ignoring risk scores, and constructs mutated sub-simulations from these segments. For instance, with $o_b = o_a = 3$ (i.e., a 6s clip), a 48s simulation yields 45 clips (skipping the first $o_b$ seconds). Although these clips often overlap, creating significant computational overhead, mutations at different start times can still produce varied outcomes. This baseline illustrates how close a technique comes to uncovering all potential failures, serving as a proxy for the ground truth in near-miss detection.

\head{\ran}\label{sec:baseline-random} 
This approach randomly selects waypoints from the original route and uses them as focal points for clipping and fuzzing. For this baseline, instead of using the misbehavior forecaster, we randomly select waypoints from the original route and apply clipping and mutation around them. 
This comparison aims to show how \tool compares against a technique that does not use any guidance to select segments for local fuzzing.

\head{\selforacle}\label{sec:baseline-selforacle}
\selforacle~\cite{2020-Stocco-ICSE} is a black-box ADS misbehavior predictor~\cite{2020-Stocco-ICSE}. Even if \selforacle was not proposed for test generation, this baseline is relevant because \selforacle is designed to detect risky situations that result in failures of ADS. 
\selforacle requires images captured by the front-facing camera for training and inference. 
We use the best configuration of \selforacle presented in the original paper, i.e., a variational autoencoder (VAE) that reconstructs driving images and uses the reconstruction loss as a measure of confidence. 
For training \selforacle, we collected 151 images at 20 FPS from the map of ``Town10'' since this map is used in all of our experiments. As the original training sets from the \ifuse and \tfuse papers are not available, we collected a training set size in line with what was described in the paper, i.e., 125k for ``Town10'' at 2 FPS. 
The autoencoder uses the Adam optimizer~\cite{Kingma2014AdamAM} to minimize the mean squared error (MSE) loss over 10 epochs, using a learning rate of 0.001. 
During inference, for each frame of the simulation, \selforacle computes a reconstruction error; we average the reconstruction errors within pairs of consecutive waypoints (i.e., a segment). This average value indicates the risk of the segment, and we rank simulation segments by this risk value. We set a threshold $\gamma=0.95$ for the expected false alarm rate in nominal conditions to identify risky conditions. 

\subsubsection{NPC Ranking}\label{sec:baseline-naive_ranking}

To evaluate the effectiveness of risk-scenario prioritization, we compare the risk-based strategy of \tool's misbehavior forecaster with a proximity ranking method. Instead of performing the complete trajectory-based critical frame ranking used in \tool~(\autoref{sec:misbehavior-forecaster}), we execute only the first step: Proximity NPC identification. This step consists of identifying all NPCs that are within a set radius during the original simulation. Thus, all NPCs in the \textit{Close NPCs} set of \tool's misbehavior forecaster are candidate risky points, which we then rank according to their order of appearance in the simulation.

\subsection{Experimental Setup}\label{sec:setup}

\subsubsection{RQ\textsubscript{1}}

We execute various configurations on the scenarios from \autoref{table:scenario-characterization} and their corresponding test cases. 
To identify NPCs in the proximity of the \ego, we used the thresholds $th_{1}=10m$ and $th_{2}=50m$ for vehicles and pedestrians, respectively. For non-crossing NPC identification, we use the $threshold=2m$. These values were tuned during preliminary pilot experiments. 

For each near miss, we clipped the test case to contain the near miss within the offsets $o_b$ and $o_a$. For simplicity, we considered offsets of the same size $o_b=o_a$. For each clipped test case, we generate $c=\mutationCount$ mutated test cases by injecting mutations according to \autoref{sec:scenario-mutator} and execute the clipped and mutated test cases.
We determine the number of risky points to analyze as follows. For each simulation, we identify risky points and consider the top $n_{rp}$ risky points from the ranked list that the misbehavior forecaster (\autoref{sec:misbehavior-forecaster}) reports, unless the number of risky points identified is less than $n_{rp}$, in which case we consider all identified risky points. We obtain the number of risky points $n_{rp}$ from the set $\{1, 2, 4\}$. Considering the length of the generated test cases, we consider $o_b = o_a \in \{3, 5\}$, thus enabling sub-simulations 6s and 10s long. 
Regarding the ranking method, we compare proximity ranking, introduced in \autoref{sec:baseline-naive_ranking}, with our Misbehavior Forecaster, which is based on risk likelihood.
For \greedy, we extract \textit{all possible} 6s and 10s clips from the original test cases by treating each second of the simulations as a risky point, i.e., taking $o_a + o_b$ seconds long cuts around each second of the original simulation. Given the high cost of this approach, we executed it only for \ifuse.

We use the number of collisions as the key performance metric for this research question (\autoref{sec:background}). We apply mutation operators on these clips, run the experiments with the same mutation count ($c$), and compute the number of failures for comparison. 

Overall, our experiment evaluates \configCount configurations of \tool (i.e., 3 risky points $\times$ 2 simulation lengths $\times$ 2 ranking methods). We empirically observed that shorter times are likely to produce invalid simulations, whereas longer times jeopardize the benefits of local fuzzing and simulation reuse. 

\subsubsection{RQ\textsubscript{2}}

We execute \ran and \selforacle with the same experimental setup from RQ\textsubscript{1}. More precisely, we use the same set of scenarios and tests (\autoref{table:scenario-characterization}), the same set of mutation operators (\autoref{sec:scenario-mutator}), and the same configurations of \tool (i.e., combinations of clip size and offsets). To assess the effectiveness of \tool's risk-based strategy, we re-run it using the most effective clip size identified in RQ\textsubscript{1}, but with only the first step of ranking (Proximity NPC identification) as described in \autoref{sec:baseline-naive_ranking}.

\subsubsection{RQ\textsubscript{3}}

We log the time that each failure was observed and report the failure rate over time as an area under the curve (AUC). In this case, the x-axis of the curve indicates time, and the y-axis of the curve indicates the cumulative number of failures observed. Intuitively, the larger the area, the better the efficiency. 

To cope with the non-determinism of the driving platform, we executed all the experiments \repetitionCount times and reported averages. 
Our dataset has a total of \totalRouteCount routes from \scenarioCount scenarios, \routesWithInfractionsIf of which have infractions when executed with \ifuse and \routesWithInfractionsTr have infractions when executed with \tfuse. Consequently, we discarded those cases, leaving \remainingRoutesIf routes (i.e., test cases) for \ifuse and \remainingRoutesTr for \tfuse. Each of these test cases is considered a seed scenario, and from each of these test cases, we derive a maximum of \maxNrp~risky points. We aimed to construct \maxNrp~clipped routes to get data for $n\_{rp} \in \{1, 2, 4\}$, as well as \mutationCount~mutated routes per clipped route from this collection of seed scenarios. Some seed scenarios do not have \maxNrp~risky points, so the number of clipped scenarios can be any value within the range $0 \leq n_{rp} \leq \maxNrp$. For \ifuse, we got a total of \totalClippedIf~clipped scenarios, and for \tfuse we got \totalClippedTr~clipped scenarios, which yields a total of \runWithRepIf~test cases for \ifuse~and \runWithRepTr~test cases for \tfuse~across 2 configurations ($(o_b+o_a) \in \{6s, 10s\}$) and \repetitionCount~repetitions.
Considering all configurations, we executed \runTotal~test cases (\numOfTechs~techniques $\times$ \mutationCount mutations $\times$ \runWithRepIf~routes for \ifuse and \numOfTechs~techniques $\times$ \mutationCount mutations $\times$ \runWithRepTr~routes for \tfuse).

It is worth noting that although test cases derived from the original input test are designed to be short running (i.e., 6s and 10s long), in practice, they tend to take longer than the estimated time because of traffic signals and vehicles getting stuck during the simulation.
In our setting, our simulations took on average \avgDurationSixs s for 6s-clips and \avgDurationTens s for 10s-clips. Thus, the average simulation time is \avgDuration~seconds, with the estimated total computing time of our experiments being more than \totalDuration~hours ($\avgDuration*\runTotal/3,600$) or around 9 days.
For the \greedy search, we created clipped and mutated sub-simulations from each second of each seed scenario (we omit the first $o_b$ seconds for each seed scenario, as it is not possible to obtain a valid simulation). In total, this process took around \greedyDuration~days of execution time across \repetitionCount repetitions with \ifuse.

\subsubsection{RQ\textsubscript{4}}

DriveFuzz is a feedback-directed test generator. We execute one iteration of DriveFuzz on the same \remainingRoutesIf~and \remainingRoutesTr~seed routes identified in RQ\textsubscript{3} for \ifuse and \tfuse respectively. We then collect the identified failing cases and conduct a second iteration of DriveFuzz on the remaining non-failing cases. The obtained \ifuse and  \tfuse routes in which DriveFuzz did not detect any failures are then used to apply the best performing configuration of \tool, identified in RQ\textsubscript{1}.
To answer this research question, a total of \totalRouteCount~full-scenarios and 480 10s-clips have been executed for each SUT, resulting in an estimated execution time of 16 hours.

\begin{table}[t]
\centering
\footnotesize
\setlength{\tabcolsep}{1.6pt}
\renewcommand{\arraystretch}{1}
\caption{RQ\textsubscript{1}: Effectiveness of \tool. Breakdown of collisions per scenario, configuration, and ADS. E=Exhaustive. \#RPs=Number of risky points. \#C.=Number of collisions.}
\label{table:rq1}
\begin{tabular}{@{}lrrrrrrrrrrrrrrr@{}}
\toprule
& \multicolumn{7}{c}{clip duration = 6s}&&\multicolumn{7}{c}{clip duration = 10s}\\
\cmidrule(r){2-8}
\cmidrule(r){9-15}

$n_{rp}$& \multicolumn{2}{c}{\cellcolor{white}1} &\multicolumn{2}{c}{2} &\multicolumn{2}{c}{\cellcolor{white}4} &\multicolumn{1}{c}{\cellcolor{white}E}&\multicolumn{2}{c}{\cellcolor{white}1} &\multicolumn{2}{c}{2} &\multicolumn{2}{c}{\cellcolor{white}4} &\multicolumn{1}{c}{\cellcolor{white}E}\\

\cmidrule(r){2-3}
\cmidrule(r){4-5}
\cmidrule(r){6-7}
\cmidrule(r){8-8}
\cmidrule(r){9-10}
\cmidrule(r){11-12}
\cmidrule(r){13-14}
\cmidrule(r){15-16}

&\#RPs &\#C. &\#RPs &\#C. &\#RPs &\#C. &\#C.&\#RPs &\#C. &\#RPs &\#C. &\#RPs &\#C. &\#C.\\

\midrule


\ifuse & & & & & & & & & & & & & & \\ [0.3em]
Scenario3 &15 &3 &30 &6 &59 &13 &51    & 15 &6 &30 &8 &59 &13 &35 \\
Scenario4 &36 &3 &68 &13 &97 &31 &128  & 35 &11 &66 &26 &95 &39 &78 \\
Scenario7 &12 &7 &26 &14 &40 &16 &66   & 11 &1 &25 &7 &41 &22 &68 \\
Scenario8 &14 &0 &30 &5 &44 &7 &18     & 13 &2 &28 &12 &42 &14 &46 \\
Scenario9 &14 &3 &25 &3 &35 &5 &7      & 14 &4 &25 &4 &37 &5 &7 \\

\cmidrule(r){2-15}

$\Sigma$ & 91 & 16 & 179 & 41 & 275 & 73 & 271 & \rpTenOne &\colTenOne &\rpTenTwo &\colTenTwo &\rpTenFour & 94 & 233 \\

\midrule

\tfuse & & & & & & & & & & & & & & \\ [0.3em]
Scenario3 &13 &1 &25 &2 &45 &2 &-  & 15 &7 &29 &11 &51 &17 &-  \\
Scenario4 &28 &2 &47 &3 &64 &6 &-  & 29 &4 &47 &9 &64 &9 &-  \\
Scenario7 &11 &0 &15 &0 &17 &0 &-  & 12 &2 &16 &2 &18 &2 &-  \\
Scenario8 &13 &0 &18 &0 &22 &0 &-  & 12 &0 &17 &0 &21 &0 &-  \\
Scenario9 &13 &0 &23 &0 &25 &0 &-  & 13 &0 &23 &0 &25 &0 &-  \\

\cmidrule(r){2-15}

$\Sigma$ &78 &3 &128 &5 &173 &8 &- & 81 &13 &132 &22 &179 &28 &- \\

\midrule

Total & 169 & 19 & 307 & 46 & 448 & 81 & - & 169 & 37 & 306 & 79 & 453 & 122 & - \\

\bottomrule
\end{tabular}
\end{table}

\subsubsection{\added{RQ\textsubscript{5}}}

\added{
We evaluate the ability of \tool\ to generalize to varying maps, agents, and weather conditions.
Specifically, we executed the best \tool configuration on all scenarios of four additional maps, namely Town01, Town02, Town03, and Town04, for both the \ifuse and \tfuse. 
Moreover, we evaluate the best-performing configuration in Town10 under an adverse weather scenario characterized by high cloudiness, fog density, wetness, and precipitation, with the sun positioned at a 45-degree angle.
To avoid confounding effects, we avoid adding further mutations (e.g., model swapping or steering angle) to the sub-simulations and measure the impact of weather on the agents.
Finally, to assess the applicability of \tool to an industrial-grade autonomous driving system, we evaluate its capability to identify risky scenarios in the \apollo\ 8 framework~\cite{apollo}.}

\subsection{Results}

\subsubsection{RQ\textsubscript{1} (effectiveness)}

\autoref{table:rq1} details the results, reporting, for different clip durations (6s or 10s), the number of risky points (\#RP), and collisions (\#Colls.) for different numbers of risky points (1, 2, or 4) and for the two evaluated SUTs (\ifuse above, \tfuse below). For comparison, the column ``\#Colls.'' under \greedy represents the maximum number of collisions detected in an exhaustive search (\autoref{sec:baseline-greedy}), only for \ifuse. 
Intuitively, as the number of risky points increases, \tool exposes more failures over time. 
For clip duration = 6s, the failure rate on \ifuse is \rateSixOne\% (\colSixOne/\rpSixOne) when one risky point is selected, \rateSixTwo\% (\colSixTwo/\rpSixTwo) when two risky points are selected, and \rateSixFour\% (\colSixFour/\rpSixFour) when four risky points are selected. 
On \tfuse, the rates are much lower, but maintain the same trend: 3.8\% (3/78), 3.9\% (5/128), and 4.6\% (8/173), respectively.
Similar considerations hold for the clip duration=10s on \ifuse where we observe failure rates of \rateTenOne\% (\colTenOne/\rpTenOne), \rateTenTwo\% (\colTenTwo/\rpTenTwo), \rateTenFour\% (\colTenFour/\rpTenFour) for the configurations with 1, 2, and 4 risky points selected, respectively. 
\tfuse achieves more modest gains: 16.0\% (13/81), 16.7\% (22/132), and 15.6\% (28/179), suggesting diminishing returns as more points are fuzzed.
The \greedy~column demonstrates the number of collisions discovered by the exhaustive search and acts as the ground truth for the potential collisions discoverable by \tool. For $o_b+o_a=6$ and $n_{rp} = 4$, \tool discovers \colSixFour~collisions compared to \greedySix~collisions discovered by \greedy, demonstrating \coverageSix \% coverage. Similarly, for $o_b+o_a=10$ and $n_{rp} = 4$, coverage for \tool is \coverageTen \% (\colTenFour x 100 / \greedyTen), demonstrating the best coverage amongst the \configCount~configurations of \tool. 

\begin{tcolorbox}[boxrule=0pt,sharp corners,boxsep=2pt,left=2pt,right=2pt,top=2.5pt,bottom=2pt]
\textbf{RQ\textsubscript{1} (effectiveness)}: \textit{
\tool exposes many collisions from near misses, with failure rates ranging between \rateSixOne-\rateTenFour\%\Space{ (\coverageSix-\coverageTen\% near-misses coverage)} for \ifuse and between 15.6-16.7 for \tfuse}. We find the configuration with clip duration=\chosenDuration{}s and number of risky points=\chosenNrp\ to provide the best trade-off in terms of failure rate.
\end{tcolorbox}

\begin{figure}[t]
\centering
\includegraphics[width=1.00\columnwidth]{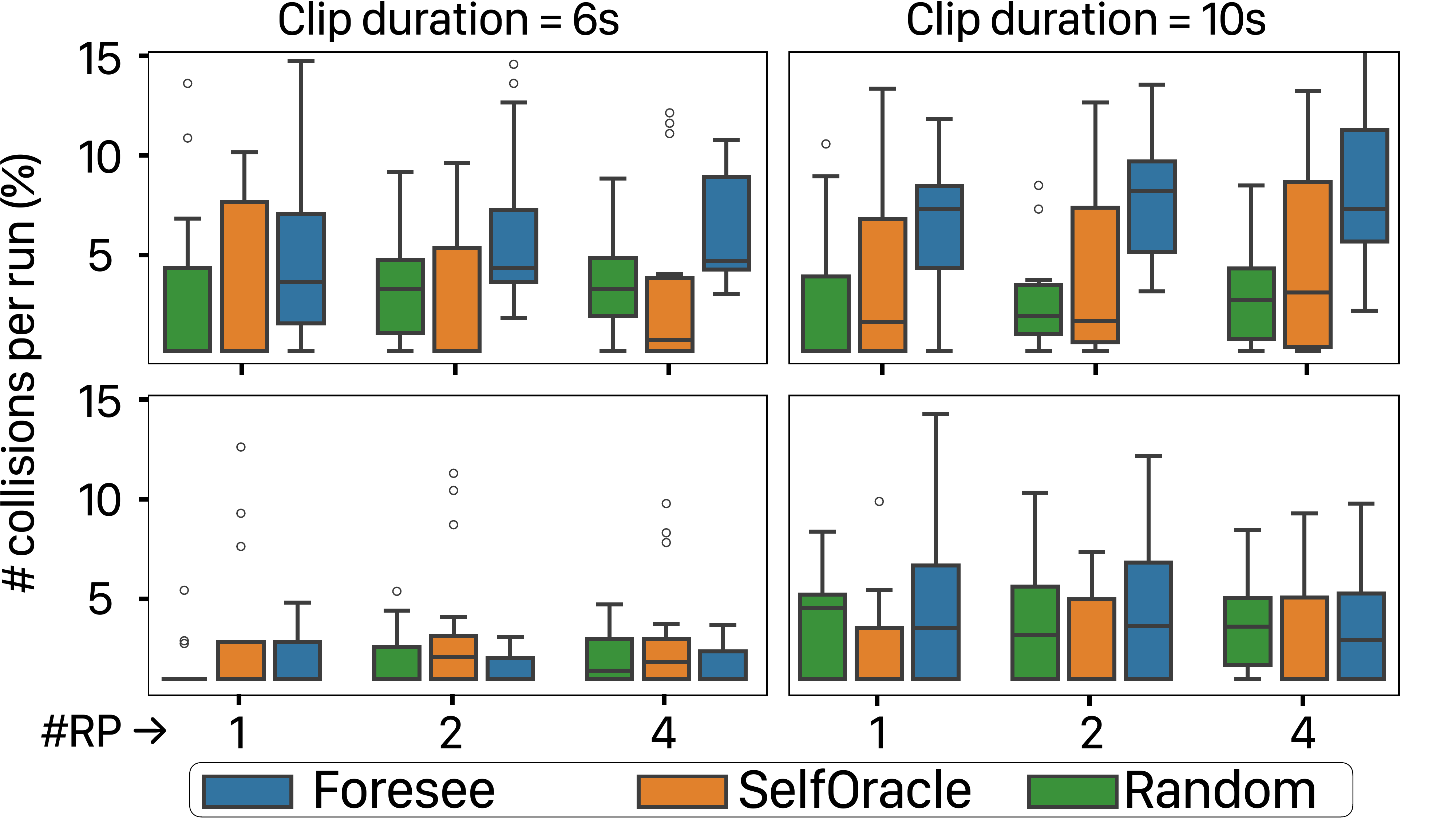}
\caption{RQ\textsubscript{2}: Comparison of different configurations.}
\label{fig:ablation_collision}
\end{figure}

\begin{table*}[t]
\centering
\footnotesize
\setlength{\tabcolsep}{5.65pt}
\renewcommand{\arraystretch}{1}
\caption{RQ\textsubscript{2}: Comparison of \ran, \selforacle, and \tool. Configuration 10 x 4 ($o_b + o_a = 10, n_{rp} = 4$).}
\label{table:rq2}
\begin{tabular}{@{}lrrrrrrrrrrrrrrrrrrrrr@{}}

\toprule

&  & \multicolumn{6}{c}{\bf \ran} & \multicolumn{6}{c}{\bf \selforacle} & \multicolumn{6}{c}{\bf \tool} \\ 

\cmidrule(r){3-8}
\cmidrule(r){9-14}
\cmidrule(r){15-20}
 
& \#TCs & F1 & F2 & F3 & F4 & F5 & $\Sigma$ & F1 & F2 & F3 & F4 & F5 & $\Sigma$ & F1 & F2 & F3 & F4 & F5 & $\Sigma$ \\

\midrule
\ifuse & & & & & & & & & & & & & & & & & & & \\ [0.3em]

Scenario3 &235 &0 &5.00 &0 &2.67 &0 &7.67 &0 &2.00 &0 &1.67 &0 &3.67 &0 &9.33 &0 &4.00 &0 &\textbf{13.33} \\
Scenario4 &390 &0.67 &0 &0 &10.00 &3.67 &14.33 &0 &0 &0 &8.33 &31.67 &\textbf{40.00} &0.67 &0 &0.67 &26.00 &11.67 &39.00 \\
Scenario7 &167 &0 &0 &0 &4.67 &0 &4.67 &0 &0 &0 &18.67 &0 &18.67 &0 &0 &1.67 &20.00 &0 &\textbf{21.67} \\
Scenario8 &179 &0.33 &0 &0.33 &4.67 &0.33 &5.67 &0 &0 &0 &1.67 &0.67 &2.33 &0 &0 &0 &14.33 &0 &\textbf{14.33} \\
Scenario9 &145 &0 &0 &0 &0.33 &0 &0.33 &0 &0 &0 &0.33 &0 &0.33 &0 &0 &0 &5.33 &0 &\textbf{5.33} \\

\cmidrule{2-20}

$\Sigma$ &1116 &1.00 &5.00 &0.33 &22.34 &4.00 &32.67 &0 &2.00 &0 &30.67 &32.34 &65.00 &0.67 &9.33 &2.34 &69.66 &11.67 &\textbf{93.66} \\

\midrule
\tfuse & & & & & & & & & & & & & & & & & & & \\ [0.3em]
Scenario3 &202 &0 &5.67 &0 &0 &0 &5.67 &0 &13.67 &0 &0 &0 &13.67 &0 &7.00 &0 &9.67 &0.33 &\textbf{17.00} \\
Scenario4 &252 &0 &0 &0 &6.67 &2.67 &\textbf{10.00} &0 &0 &0 &8.33 &1.33 &9.67 &0.33 &0 &0 &8.33 &0.33 &9.00 \\
Scenario7 &71 &0 &0 &0 &1.33 &0 &1.33 &0 &0 &0 &0 &0 &0 &0 &0 &0 &2.00 &0 &\textbf{2.00} \\
Scenario8 &78 &0.33 &0 &0 &3.33 &0 &\textbf{3.67} &0 &0 &0 &0 &0 &0 &0 &0 &0 &0.33 &0 &0.33 \\
Scenario9 &92 &0 &0 &0 &0 &0 &0 &0 &0 &0 &0 &0 &0 &0 &0 &0 &0 &0 &0 \\

\cmidrule{2-20}
$\Sigma$ & 695 &0.33 &5.67 &0 &11.33 &2.67 &20.67&0 &13.67 &0 &8.33 &1.33 &23.34 &0.33 &7.00 &0 &20.33 &0.66 &\textbf{28.33} \\
\midrule
Total & 1811 & 1.33 & 10.67 & 0.33 & 33.67 & 6.67 & 53.34 & 0 & 15.67 & 0 & 39.00 & 33.67 & 88.34 & 1.00 & 25.33 & 2.34 & 89.99 & 12.33 & 121.99\\
\bottomrule
\end{tabular}
\end{table*}

\subsubsection{RQ\textsubscript{2} (comparison)}

\autoref{fig:ablation_collision} shows the distributions of the number of collisions for \tool, \selforacle, and \ran with \ifuse~(top) and \tfuse~(bottom). The left and right figures show the results for clip durations of 6s and 10s, respectively. \tool outperforms the baselines in all configurations except clip duration=6s for \tfuse~ where no technique achieves adequate performance. We measured the statistical significance of the differences using the nonparametric \added{Wilcoxon rank-sum test}~\cite{Wilcoxon1945}, with $\alpha=0.05$, and the magnitude of the differences using Cohen's effect size $d$~\cite{cohen1988statistical}. To account for multiple comparisons, we also applied Holm–Bonferroni correction~\cite{abdi2010holm} to our \textit{p}-values. 

For most configurations with \ifuse, the differences between \tool and both baselines are statistically significant (6x2, 6x4, 10x2, and 10x4), i.e., the $p$ value $<$ 0.05 with a medium to large effect size. For some configurations (6x1 and 10x1), only the differences between \tool and \ran are statistically significant, with medium/large effect sizes. \changed{The effectiveness of Foresee seems to be autopilot-dependent, as expected. InterFuser is a more stable agent, and configurations with richer sampling (e.g., 10×2, 10×4, 6×4, 10x4) yield higher effect sizes, indicating that predictive guidance benefits from more informative search spaces. TransFuser proved more failure-prone, and thus simpler configurations (e.g., 10×1, 10×2) are sufficient.}

For \tfuse, configurations with clip duration=6s produce a negligible number of collisions with any technique, indicating this configuration is not ideal for testing \tfuse. For configurations with clip duration=10s, \tool~has an advantage over \selforacle~with medium effect sizes for configurations 10x1 and 10x2, as well as a small effect size for 10x4. However, \tool~has only a small effect size over \ran~in configuration 10x1, 10x2, with a very small effect size (Cohen's $d$=0.11) in configuration 10x4.

We provide detailed comparisons for the configuration \chosenDuration{}x\chosenNrp, which is the one with the best failure rate from RQ\textsubscript{1} for \ifuse. For \tfuse, failure rates across all configurations with clip duration=\chosenDuration{} are very similar, so we chose \chosenDuration{}x\chosenNrp{} for consistency.
\autoref{table:rq2} shows, for each technique (rows) and SUT (\ifuse above, \tfuse below), the average number of failures and their nature. 
Column \#TCs shows the number of test cases associated with a given scenario. 
Columns F1, F2, F3, F4, and F5 show the different kinds of collisions detected, respectively related to collisions involving the \ego with elements beyond the road, such as pavements or poles (F1), pedestrians (F2), and frontal (F3), lateral (F4), or rear (F5) collisions with other vehicles. Column $\Sigma$ shows totals. Results are presented for each scenario separately, as well as an aggregate. 
The table reinforces the effectiveness of \tool over the baselines, across all scenarios, failure types, and SUTs. Overall, \tool achieves, for \ifuse, a failure rate increase of +\FOImprovementOverRandomIfuse\% and +\FOImprovementOverSOIfuse\% with respect to \ran and \selforacle. 

For \tfuse, the failure rate increases were +\FOImprovementOverRandomTfuse\% and +\FOImprovementOverSOTfuse\% over \ran and \selforacle, respectively. Over the two SUTs, \tool~achieves +\FOImprovementOverRandom\% and +\FOImprovementOverSO\% failure rate compared to \ran~and \selforacle. It also achieves a higher diversity of failure kinds observed, even though failure {F4 (lateral collisions with other vehicles) is the most prevalent. 

\begin{table}[t]
\centering
\footnotesize
\setlength{\tabcolsep}{2.65pt}
\renewcommand{\arraystretch}{1}
\caption{RQ\textsubscript{2}: Effectiveness of \tool risk-based ranking. \#RPs=Number of risky points. \#C. = Number of collisions.} 
\label{table:ablation}
\begin{tabular}{@{}lccccccccccccc@{}}
\toprule
&\multicolumn{6}{c}{\bf Proximity rank} & \multicolumn{6}{c}{\bf Risk-based rank} \\
\cmidrule(r){2-7}
\cmidrule(r){8-13}
$n_{rp}$&\multicolumn{2}{c}{1} &\multicolumn{2}{c}{2} &\multicolumn{2}{c}{4}&\multicolumn{2}{c}{1} &\multicolumn{2}{c}{2} &\multicolumn{2}{c}{4}\\
\cmidrule(r){2-3}
\cmidrule(r){4-5}
\cmidrule(r){6-7}
\cmidrule(r){8-9}
\cmidrule(r){10-11}
\cmidrule(r){12-13}
&\#RPs &\#C. &\#RPs &\#C. &\#RPs &\#C. &\#RPs &\#C. &\#RPs &\#C. &\#RPs &\#C.\\

\midrule
\ifuse & & && & &\\ [0.3em]
Scenario3 &15 &0 &29 &1 &58 &7  & 15 &6 &30 &8 &59 &13 \\
Scenario4 &37 &1 &73 &6 &131 &8 & 35 &11 &66 &26 &95 &39 \\
Scenario7 &11 &2 &23 &3 &40 &3  & 11 &1 &25 &7 &41 &22\\
Scenario8 &17 &2 &33 &4 &62 &7  & 13 &2 &28 &12 &42 &14\\
Scenario9 &10 &4 &23 &5 &44 &5  & 14 &4 &25 &4 &37 &5 \\
\midrule
$\Sigma$ &90 &9 &181 &19 &335 &30 & \rpTenOne &\colTenOne &\rpTenTwo &\colTenTwo &\rpTenFour &93 \\
\cmidrule(r){2-3}
\cmidrule(r){4-5}
\cmidrule(r){6-7}
\cmidrule(r){8-9}
\cmidrule(r){10-11}
\cmidrule(r){12-13}
\#C / \#RP &\multicolumn{2}{c}{0.10} &\multicolumn{2}{c}{0.10} &\multicolumn{2}{c}{0.08} &\multicolumn{2}{c}{0.27} & \multicolumn{2}{c}{0.32} & \multicolumn{2}{c}{\textbf{0.33}}\\
\midrule
\tfuse & & && & &\\ [0.3em]
Scenario3 &15 &5 &29 &8 &55 &13  & 15 &7 &29 &11 &51 &17 \\
Scenario4 &39 &2 &76 &2 &141 &3  & 29 &4 &47 &9 &64 &9 \\
Scenario7 &15 &0 &29 &1 &57 &6   & 12 &2 &16 &2 &18 &2 \\
Scenario8 &18 &2 &36 &3 &67 &7   & 12 &0 &17 &0 &21 &0 \\
Scenario9 &17 &3 &33 &3 &64 &5   & 13 &0 &23 &0 &25 &0 \\
\midrule
$\Sigma$ &104 &12 &203 &17 &384 &34 &     81 &13 &132 &22 &179 &28 \\

\cmidrule(r){2-3}
\cmidrule(r){4-5}
\cmidrule(r){6-7}
\cmidrule(r){8-9}
\cmidrule(r){10-11}
\cmidrule(r){12-13}
\#C / \#RP &\multicolumn{2}{c}{0.12} &\multicolumn{2}{c}{0.08} &\multicolumn{2}{c}{0.09} &\multicolumn{2}{c}{0.16} & \multicolumn{2}{c}{\textbf{0.17}} & \multicolumn{2}{c}{0.16} \\
\bottomrule
\end{tabular}
\end{table}

To assess the contribution of \tool's full risk-based ranking strategy, \autoref{table:ablation} reports the results for both the Proximity-based NPC selection baseline (Proximity rank, left-hand side) and \tool's complete ranking method (\tool, right-hand side). The table shows the number of risky points evaluated and the number of collisions induced across scenarios, for different \#RP values.

Across both systems under test, \tool's ranking strategy consistently improves efficiency, measured as collisions per risky point, compared to the proximity-only baseline. Under \ifuse, \tool produces both more total collisions and higher efficiency. For example, with \#RP = 2, \tool triggers \changed{57} collisions from 174 risky points (\changed{0.33} collisions per RP), while the proximity baseline triggers just 19 from 181 points (0.10 per RP). With \#RP = 4, the difference is even more pronounced: \changed{93} collisions for \tool (\changed{0.34} per RP) versus 30 for proximity (0.09 per RP).
On \tfuse, the trend is more nuanced. The proximity baseline reaches slightly more total collisions (e.g., 34 vs. 28 at \#RP = 4), but it requires significantly more exploration, 384 risky points compared to 179. This results in lower efficiency (0.09 collisions per RP vs. 0.16 for \tool). Thus, while the proximity strategy occasionally triggers more failures in absolute terms, \tool achieves comparable or better outcomes with far fewer test executions, making it more cost-effective.

\begin{tcolorbox}[boxrule=0pt,sharp corners,boxsep=2pt,left=2pt,right=2pt,top=2.5pt,bottom=2pt]
\textbf{RQ\textsubscript{2} (comparison)}: \textit{
\tool outperforms the considered baselines, with a failure rate increase of +\FOImprovementOverRandom\% and +\FOImprovementOverSO\% with respect to \ran and \selforacle, respectively. Ablation results confirm that \tool risk-based ranking improves failure discovery efficiency over a proximity-only baseline, across both systems under test, highlighting the benefit of \tool risk-based prioritization.
}
\end{tcolorbox}

\begin{figure}[t]
    \centering
    \includegraphics[trim={0cm 11.7cm 0cm 0.0cm}, clip, width=\columnwidth]{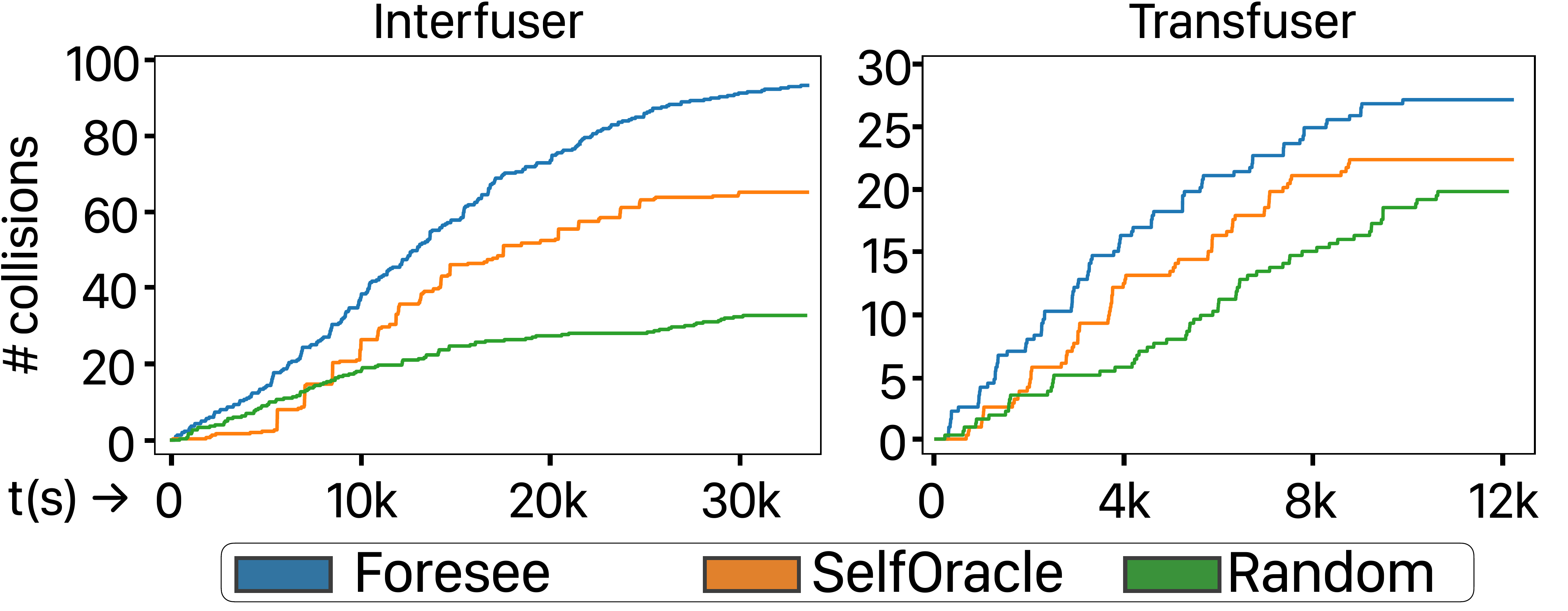}
    \caption{RQ\textsubscript{3}: Efficiency of \ran, \selforacle, and \tool for configuration 10 x 4 ($o_b+o_a=10$ and $n_{rp}=4$).}
    \label{fig:auc_10s}
\end{figure}

\subsubsection{RQ\textsubscript{3} (efficiency)}

The \greedy search, executed on \ifuse, showed the maximum number of collisions discoverable, but \tool is more efficient in finding the collisions. For $o_b+o_a=6s$ and $n_{rp}=4$, \greedy exposed \greedySixColByHour~collisions per hour ($\greedySix / \greedySixSimTime$) compared to \foreseeSixColByHour~collisions per hour ($\colSixFour / \avgDurationSix$) by \tool, resulting in a \efficiencySixPercent \% efficiency increase. Similarly, for $o_b+o_a=10s$ and $n_{rp}=4$, \greedy~discovered \greedyTenColByHour~collisions per hour ($\greedyTen / \greedyTenSimTime$) compared to \foreseeTenColByHour~collisions per hour ($\colTenFour / \avgDurationTen$) discovered by \tool, a \efficiencyTenPercent \% increase.

\autoref{fig:auc_10s} shows, for both \ifuse and \tfuse, the cumulative number of collisions detected by the techniques over time and the area under the curve (AUC) associated with the corresponding plots. 
The result indicates the superior ability of \tool over the baselines to efficiently expose failures, as evidenced by the position of \tool's plot relative to the plot of the other techniques (and higher AUC score). \tool has an AUC score of \foreseeAUCIf, which is \foreseeBySelfOracleAUCIf higher than the AUC score of \selforacle and \foreseeByRandomAUCIf higher than the AUC score of \ran for \ifuse, and shows a similar trend for \tfuse (AUC=\foreseeAUCTr, an increase of \foreseeBySelfOracleAUCTr~and \foreseeByRandomAUCTr~against \selforacle and \ran). Across both SUTs, \tool~is \foreseeBySelfOracleAUC~faster than \selforacle~and \foreseeByRandomAUC~faster than \ran.

\begin{tcolorbox}[boxrule=0pt,sharp corners,boxsep=2pt,left=2pt,right=2pt,top=2.5pt,bottom=2pt]
\textbf{RQ\textsubscript{3} (efficiency)}: \textit{
\tool identifies collisions significantly faster than the baselines. For \ifuse, the AUC of \tool is \foreseeBySelfOracleAUCIf~higher than \selforacle's and \foreseeByRandomAUCIf~higher than \ran's, while for \tfuse, AUC increases by \foreseeBySelfOracleAUCTr~and \foreseeByRandomAUCTr~compared to \selforacle~and \ran~(\foreseeBySelfOracleAUC~and \foreseeByRandomAUC~faster overall).
}
\end{tcolorbox}

\subsubsection{RQ\textsubscript{4}: Complementarity}

Table~\ref{table:rq4} presents the number of infractions detected in the evaluated scenarios over two iterations of DriveFuzz, together with the additional failures identified by \tool when applied to nonfailing cases. Each row reports the plain infractions (on the original unfuzzed routes), followed by the results of DriveFuzz and \tool, for each iteration. The data illustrate how \tool uncovers additional failures beyond those detected by DriveFuzz alone.

For \ifuse, in Iteration~1, DriveFuzz discovered 18 infractions, while \tool found 19 additional ones. In Iteration~2, DriveFuzz added 15 more, and \tool contributed 12 additional failures. This increased the total failure count from 17 (baseline) to 81, corresponding to a 376.47\% increase. \tool alone uncovered 31 failures, compared to 33 by DriveFuzz, which is a 93.94\% increase over DriveFuzz's findings.

For \tfuse, DriveFuzz uncovered 21 infractions in Iteration~1 and 16 in Iteration~2, while \tool identified 11 and 6 additional failures, respectively. The cumulative failure count thus rose from 9 to 63, a 600.00\% increase over the baseline. \tool contributed 17 failures in total, an increase of 45.95\% compared to the 37 found by DriveFuzz. Notably, in Scenarios~7 and~8 for \tfuse, which had zero baseline infractions, Drivefuzz and \tool combined revealed multiple failures (\changed{10} and \changed{9})}.

\begin{table}[t]
\centering
\footnotesize
\setlength{\tabcolsep}{2pt}
\renewcommand{\arraystretch}{1}
\caption{RQ\textsubscript{4}: Effectiveness of \tool with DriveFuzz. \#F=Number of failures.  +\%=Percentage increase w.r.t. Plain.}
\label{table:rq4}
\begin{tabular}{@{}lrrrrrrrrr@{}}
\toprule   
&  & \multicolumn{4}{c}{\bf Iteration 1} & \multicolumn{4}{c}{\bf Iteration 2} \\ 
\cmidrule(r){3-6}
\cmidrule(r){7-10}
& Plain & DriveFuzz & \tool & \multicolumn{2}{c}{$\sum$} & DriveFuzz  & \tool &  \multicolumn{2}{c}{$\sum$}\\
\cmidrule(r){2-2}
\cmidrule(r){3-3}
\cmidrule(r){4-4}
\cmidrule(r){5-6}
\cmidrule(r){7-7}
\cmidrule(r){8-8}
\cmidrule(r){9-10}
& \#F & +\#F & +\#F & \#F & +\% & +\#F & +\#F & \#F & +\%\\
\midrule
\ifuse\\
Scenario 3 &1 &3 &2 &6 &500 &2  &1 &9 &800\\
Scenario 4 &6 &4 &5 &15 &150 &7  &5 &27 &350\\
Scenario 7 &5 &3 &3 &11 &120 &0  &1 &12 &140\\
Scenario 8 &2 &6 &7 &15 &650 &3  &5 &23 &1050\\
Scenario 9 &3 &2 &2 &7 &133 &3 &0 &10 &233\\ 
\cmidrule(r){2-10}
$\Sigma$ &17 &18 &19 &54 & 217.65 &15 &12 &81 &376.47\\ 

\midrule
\tfuse \\
Scenario 3 &1 &4 &4 &8 &700 &8  &1 &18 &1700\\
Scenario 4 &5 &9 &1 &10 &100 &5  &1 &21 &320\\
Scenario 7 &0 &4 &2 &6 &- &0  &4 &10 &-\\
Scenario 8 &0 &3 &4 &7 &- &2  &0 &9 &-\\
Scenario 9 &3 &1 &0 &4 &33 &1 &0 &5 &66.67\\ 
\cmidrule(r){2-10}
$\Sigma$ &9 &21 &11 &41 &355.56 &16 &6 &63 &600\\
\bottomrule
\end{tabular}
\end{table}

\begin{tcolorbox}[boxrule=0pt,sharp corners,boxsep=2pt,left=2pt,right=2pt,top=2.5pt,bottom=2pt]
\textbf{RQ\textsubscript{4} (complementarity)}: \textit{
\tool complements the state-of-the-art fuzzer DriveFuzz by uncovering 31 additional failures for \ifuse and 17 for \tfuse, corresponding to a 93.94\% and 45.95\% increase over the failures found by DriveFuzz alone. In multiple scenarios (e.g., Scenario~7 and~8 in \tfuse), \tool detected failures where DriveFuzz found none, suggesting its effectiveness in extending the coverage of existing fuzzing campaigns.
}
\end{tcolorbox}

\subsubsection{\added{RQ\textsubscript{5}: Generalizability}}\label{sec:rq5}

\added{
\autoref{table:rq5} shows \tool's capability of revealing infractions on four additional maps: Town01, Town02, To,wn03 and Town04 with the \ifuse\ and \tfuse agents. \tool\ can effectively discover infractions of different categories across all towns and agents. We find that the most common infraction type is F4 (lateral collision), observed 181 times, whereas F3 (frontal collision) is the most uncommon, observed 10 times. For \tfuse, we find 115 previously unknown collisions across 1,079 test cases. For \ifuse, we find 176 previously unknown collisions across 1,307 test cases.}

\added{
\autoref{table:rq5-apollo} show \tool's effectiveness with the \apollo~\cite{apollo} framework. Unlike \ifuse\ and \tfuse\ which are end-to-end systems, \apollo\ is a multi-modal system~\cite{xiao2020multimodal} where different modules (e.g., planner, control, PID) control different aspects of driving. Consequently, its behavior differs from that of the other agents, as reflected in the distribution of collision types in \autoref{table:rq5-apollo}.
Aside from a few lateral collisions and instances where the vehicle remained stationary, most collisions were classified as type F1 (object). We discuss these failures in more detail in \autoref{sec:discussion}.
}

\added{
\autoref{table:rq5-weather} demonstrates \tool's ability to reveal infractions under heavy weather conditions. In these conditions, both \tfuse\ and \ifuse\ reveal 34 lateral collisions (F4), which is the majority. It also reveals 8 pedestrian (F2) and 4 rear (F4) collisions and avoids frontal/object collisions. In heavy weather, both agents drive more carefully and slowly to avoid frontal collisions and colliding with roadside collisions, but the effects of low visibility causes lateral and pedestrian collisions, while driving too slow in risky situations causes some rear collisions.
}

\begin{tcolorbox}[boxrule=0pt,sharp corners,boxsep=2pt,left=2pt,right=2pt,top=2.5pt,bottom=2pt]
\textbf{RQ\textsubscript{5} (generalizability)}: 
\added{\textit{\tool\ is capable of generalizing under different maps, weather conditions, and an industrial-grade multi-modal agent, \apollo.
For example, \tool\ reveals 291 collisions from previously failure-free scenarios in four different towns with \tfuse\ and \ifuse, 30 new collisions with \apollo, and 46 new collisions under heavy weather conditions.}}
\end{tcolorbox}

\begin{table}[t]
    \centering
    \tiny
    \caption{\added{RQ\textsubscript{5}: Generalizability. Impact of town, agent (Apollo), and heavy weather. Configuration 10 x 4 ($o_b + o_a = 10, n_{rp} = 4$).}}
    \begin{subtable}{0.5\textwidth}
        \centering
\caption{\label{table:rq5}Towns}
\resizebox{0.9\columnwidth}{!}{
\begin{tabular}{llrrrrrrrr}\toprule
Map &Agent &\#TCs &F1 &F2 &F3 &F4 &F5 & Stuck &$\Sigma$ \\\midrule
\multirow{2}{*}{Town01} &\ifuse &238 &8 &3 &0 &17 &4 &0&32 \\
 &\tfuse &205 &0 &0 &0 &10 &6 &0&16 \\\midrule
\multirow{2}{*}{Town02} &\ifuse &160 &22 &0 &5 &17 &3 &0&47 \\
 &\tfuse &143 &0 &0 &0 &1 &0 &0&1 \\\midrule
\multirow{2}{*}{Town03} &\ifuse &512 &0 &0 &1 &25 &0 &0&27 \\
 &\tfuse &324 &0 &15 &0 &15 &0 &0&30 \\\midrule
\multirow{2}{*}{Town04} &\ifuse &397 &24 &0 &0 &44 &1 &0&70 \\
 &\tfuse &407 &0 &0 &4 &52 &12 &0&68 \\\midrule
 $\Sigma$ & &2386	&54	&18	&10	&181	&26 &0 &291\\
\bottomrule
\end{tabular}}        
    \end{subtable}
    \hfill
    \begin{subtable}{0.5\textwidth}
        \centering
        \small
\caption{\label{table:rq5-apollo}Apollo}
\begin{tabular}{lrrrrrrrr}\toprule
Map &\#TCs &F1 &F2 &F3 &F4 &F5 & Stuck &$\Sigma$ \\\midrule
Town01 & 80 & 15 & 0 & 0 & 1 & 0 &0 & 16 \\
Town03 & 64 & 6 & 0 & 0 & 3 & 0 & 5 & 14 \\\midrule
$\Sigma$ & 144 & 21 & 0 & 0 & 4 & 0 & 5 & 30 \\
\bottomrule
\end{tabular}
    \end{subtable}
    \begin{subtable}{0.5\textwidth}
    \small
        \centering
\caption{\label{table:rq5-weather}Heavy weather}
\begin{tabular}{lrrrrrrrr}\toprule
Agent &\#TCs &F1 &F2 &F3 &F4 &F5 & Stuck &$\Sigma$ \\\midrule
\ifuse &539 &0 &3 &0 &29 &4 &0 &36 \\
\tfuse &237 &0 &5 &0 &5 &0 &0 &10 \\\midrule
$\Sigma$ &776 &0 &8 &0 &34 &4 &0 &46 \\
\bottomrule
\end{tabular}
    \end{subtable}
\end{table}

\subsection{Qualitative Analysis}\label{sec:discussion}

To evaluate whether \tool exploits segments already close to failure or uncovers meaningful, non-trivial vulnerabilities, we conducted a qualitative analysis on the data from the exhaustive search (RQ\textsubscript{1}). For each failure-free test, we found the most critical ego–NPC interactions using simulation logs and characterized them using three interpretable features: minimum time-to-collision (TTC), relative distance, and relative speed.
We labeled each interaction based on whether it led to a failure during fuzzing and trained a Random Forest classifier~\cite{ho1995random} on each scenario to predict failure-prone interactions. The models achieved recall in the 0.64-0.86 range~(except for Scenario 9 at 0.31), and precision in the 0.12-0.41 range. Results suggest that, while failures can often be detected, the boundary between safe and unsafe behaviors is complex.

Failures rarely stemmed from a single extreme factor (e.g., low TTC), but instead from nuanced combinations. In Scenario 3, failures occurred with high TTC ($>$1482s) and long distances ($>$150m) when paired with modest speeds ($>$$-$3.1m/s). Scenario 4 showed failures at distances up to 83.99m when relative speed was mild ($>$$-$4.2m/s). In Scenario 8, short distances ($\leq$3.29m) combined with low relative speeds ($>$0.0m/s) triggered failures. Scenario 9 revealed failures even with high TTC ($>$30s). 

\added{
Regarding \apollo's failures, the driving agent generally performs well in avoiding other vehicles under nominal conditions. However, when a nearby NPC vehicle encroaches on the ego lane, the local planner tends to generate sharper avoidance trajectories. This behavior leads the control module to produce oscillatory, wobble-like movements, resembling the response of a PID controller under suboptimal tuning.
These oscillations can amplify over time, causing the vehicle to drift off the lane and eventually collide with static objects such as walls, poles, or benches—obstacles that the perception system does not consistently detect in advance. Consequently, most failures correspond to object collisions (type F1). Under normal driving conditions, these crashes do not occur; the revealed failures highlight a limited ability of \apollo's planning and control modules to generalize to such corner cases.
}

\subsection{Threats to Validity}\label{sec:ttv}

\head{Internal Validity}
All variants of \tool, \selforacle, and random were evaluated under identical experimental conditions and the same test set. The main threat concerns the correctness of our test script implementations, which we verified thoroughly. We could not train \selforacle using the same dataset as \ifuse, nor use pre-trained models due to simulator differences. We mitigated this by training the VAE following standard guidelines~\cite{2020-Stocco-ICSE,2021-Stocco-JSEP}, and our version of \selforacle performs competitively.

Regarding the ADS, which could bias results if unsuitable for driving, we used two top-ranking, publicly available models from the CARLA leaderboard. As for the simulation platform, we adopted CARLA, widely used in recent failure prediction research~\cite{morlot}.

One threat relates to the quality of input test cases, specifically, whether scenarios already include near-misses, making failures easier to induce. To assess this, we analyzed whether failures found by \tool stemmed from obviously risky inputs. As discussed in \autoref{sec:discussion}, decision tree analyses show that failures arise from subtle, multi-feature conditions rather than extreme TTC, distance, or speed values, suggesting \tool reveals non-trivial vulnerabilities.

Another concern involves the realism of scenario mutations. Unlike global mutation tools like DriveFuzz~\cite{drivefuzz}, our method applies minimal, targeted modifications to existing simulations, preserving NPC count and adjusting only those closest to the ego vehicle. We use conservative perturbations, such as slight steering changes or vehicle swaps within the same class, aligned with localization error thresholds~\cite{localization_req} and established targeted testing techniques~\cite{boundary}.
\changed{Limited repetition of our experiments may affect the robustness of our results, as simulation-based testing is computationally expensive.}

\head{External Validity}
The limited number of self-driving systems in our evaluation constitutes a threat to the generalizability of our results to other ADS. Results may not generalize when considering other simulation platforms or ADSs. 

\section{Related Work}\label{sec:related-work}

\noindent\textbf{Test Generation for Autonomous Driving.}
The majority of test generation techniques employ search-based techniques for DNN-based ADS testing~\cite{arxiv.2203.12026,2020-Riccio-FSE,Abdessalem-ASE18-1,Abdessalem-ASE18-2,Abdessalem-ICSE18,deepxplore,deeptest,deeproad,Kim:2019:GDL:3339505.3339634}. 
Gambi et al.~\cite{Gambi:2019:ATS:3293882.3330566} propose search-based test generation for ADS based on procedural content generation.
DriveFuzz uses the physical states of the vehicle and oracles based on real-world traffic rules to guide the fuzzer towards finding misbehaviors~\cite{drivefuzz}.
AutoFuzz~\cite{zhongETAL2021} focuses on fuzzing the test scenario specification. Before fuzzing, it uses a seed selection mechanism based on a binary classifier that selects likely traffic-violating seeds.
AV-Fuzzer~\cite{liETAL2020} uses a genetic algorithm that is informed by the positioning of globally monitored NPCs in each scenario in the driving environment. 
Cheng et al.~\cite{10.1145/3597926.3598072} propose BehaviorMiner, an unsupervised model that extracts the temporal features from certain given scenarios and performs a clustering-based abstraction to group behaviors with similar features into abstract states. 
Neelofar and Aleti~\cite{NeelofarTOSEM} propose characterizing critical scenarios for ADS using a combination of static and dynamic features. 
SimADFuzz~\cite{yang2024simadfuzzsimulationfeedbackfuzztesting} uses distance-guided mutation strategies to enhance the probability of interactions among vehicles in generated scenarios. 
\changed{Koren et al.~\cite{Koren} and Corso et al.~\cite{Corso} use Monte Carlo tree search and deep reinforcement learning to find collision scenarios.
Similarly, GARL~\cite{garl} uses reinforcement learning to detect violations in marker-based autonomous landing systems.
}

Test generators are designed to maximize the number of failures and consider \textit{whole} test cases. While the exploration is guided towards critical regions, the search budget is consumed by running test cases that do not result in failing conditions. 
Our approach forecasts potential \ego states to predict infractions with NPCs and focuses on local segments within test cases. 

\noindent\textbf{Focused Test Generation for Autonomous Driving.}
Although focused test case generation has been a subject of extensive study and application in the context of software testing, its application to ADS is a new field that has been underexplored. To the best of our knowledge, only two approaches have been proposed. DeepAtash-LR by Zohdinasab et al.~\cite{10.1145/3664605} improves targeted test generation by using a surrogate model to avoid executing whole test cases. In contrast, we focus on reusing simulation segments.
TM-Fuzzer~\cite{tmfuzzer} dynamically generates NPCs in the neighborhood of the ego vehicle to increase the likelihood of failures. Differently, we keep the number of NPCs the same and introduce only a minimal modification by changing their type, which impacts their kinematic behavior. Also, our framework promotes the reuse of simulation resources, which is not the case with any of the existing approaches.

\head{Anomaly Detection in Autonomous Driving}
We already discussed SelfOracle~\cite{2020-Stocco-ICSE}, for which we performed an explicit empirical comparison in this work. Similarly, 
DeepGuard~\cite{10.1007/s10515-021-00310-0} uses the reconstruction error by VAEs to prevent collisions of vehicles with the roadside. 
ThirdEye~\cite{2022-Stocco-ASE} uses the attention maps from the explainable AI domain to predict misbehaviors of self-driving cars.
Other works~\cite{2021-Stocco-JSEP,2020-Stocco-GAUSS} use continual learning to minimize the false positives of a black-box failure predictor, whereas other researchers use uncertainty quantification~\cite{2024-Grewal-ICST} or probabilistic time-series~\cite{2025-Sharifi-TOSEM}.

Our approach differs from the aforementioned approaches because it uses a risky score of the system synthesized from a forecasting mechanism of the \ego kinetics. 

\section{Conclusions}\label{sec:conclusions}

Simulation-based testing is highly useful in autonomous vehicle testing, but the cost of revealing faults relative to the time to run simulations is very high. 
We propose \tool, an approach to optimize simulation-based testing by reusing segments of simulations that produce near-failing situations. Our approach uses a custom misbehavior forecaster to detect near misses and fuzz the state of the simulation locally to produce failures quickly. 
Experimental results show that failure-free scenarios embed many near-failing situations that \tool can accurately detect, and that many of these cases result in failures when minimal perturbations are introduced. 
\tool provides initial yet strong evidence that guiding fuzzing with misbehavior forecasting is a promising approach to uncovering hidden failures in ADS. 
In the future, we plan to evaluate additional forecasting methods based on multi-horizon and multi-variate time series.

\begin{acks}
This work was partially supported by the National Science Foundation (NSF) under Grant Nos. CCF-2349961 and CCF-2319472. Lambertenghi and Stocco were supported by the Bavarian Ministry of Economic Affairs, Regional Development, and Energy.
\end{acks}

\balance
\bibliographystyle{ACM-Reference-Format}
\bibliography{ref}

\end{document}